\documentclass[12pt,a4paper]{article}
\usepackage{graphicx}
\usepackage{subeqnarray}
\begin{document}
%
%
%
%
\newenvironment{lefteqnarray}{\arraycolsep=0pt\begin{eqnarray}}
{\end{eqnarray}\protect\aftergroup\ignorespaces}
\newenvironment{lefteqnarray*}{\arraycolsep=0pt\begin{eqnarray*}}
{\end{eqnarray*}\protect\aftergroup\ignorespaces}
\newenvironment{leftsubeqnarray}{\arraycolsep=0pt\begin{subeqnarray}}
{\end{subeqnarray}\protect\aftergroup\ignorespaces}
\newcommand{\diff}{{\rm\,d}}
\newcommand{\pprime}{{\prime\prime}}
\newcommand{\szeta}{\mskip 3mu /\mskip-10mu \zeta}
\newcommand{\srho}{\mskip 3mu /\mskip-10mu \rho}
\newcommand{\sr}{\mskip 3mu /\mskip-9mu r}
\newcommand{\sR}{\mskip 3mu /\mskip-11mu R}
\newcommand{\sV}{\mskip 3mu /\mskip-10mu V}
\newcommand{\sP}{\mskip 3mu /\mskip-10mu p}
\newcommand{\sPM}{\mskip 3mu /\mskip-11mu P}
\newcommand{\sT}{\mskip 3mu /\mskip-9mu T}
\newcommand{\FC}{\mskip 0mu {\rm F}\mskip-10mu{\rm C}}
\newcommand{\appleq}{\stackrel{<}{\sim}}
\newcommand{\appgeq}{\stackrel{>}{\sim}}
\newcommand{\quadr}{\overline\sqcup}
\newcommand{\Int}{\mathop{\rm Int}\nolimits}
\newcommand{\Nint}{\mathop{\rm Nint}\nolimits}
\newcommand{\arcsinh}{\mathop{\rm arcsinh}\nolimits}
\newcommand{\range}{{\rm -}}
\newcommand{\displayfrac}[2]{\frac{\displaystyle #1}{\displaystyle #2}}
%
%
\title{The intrinsic beauty of polytropic spheres in reduced variables}

\author{
 {R.~Caimmi}\footnote{
{\it Physics and Astronomy Department, Padua University,
Vicolo Osservatorio 3/2, I-35122 Padova, Italy.   Affialiated up to September
30th 2014.  Current status: Studioso Senior.   Current position: in retirement
due to age limits.}\hspace{50mm}
email: roberto.caimmi@unipd.it~~~
fax: 39-049-8278212}
\phantom{agga}}

\maketitle
\begin{quotation}
\section*{}
\begin{Large}
\begin{center}

 Abstract

\end{center}
\end{Large}
\begin{small}

\noindent\noindent
The concept of reduced variables is revisited with regard to van der Waals'
theory and an application is made to polytropic spheres, where the reduced
radial coordinate is $\sr=r/R=\xi/\Xi$, $R$ radius, and the reduced density is
$\srho=\rho/\lambda=\theta^n$, $\lambda$ central density.   Reduced density
profiles are plotted for several polytropic indexes within the range,
$0\le n\le5$, disclosing two noticeable features.   First, any point of
coordinates, $(\sr,\srho)$, $0\le\sr\le1$, $0\le\srho\le1$, belongs to a
reduced density profile of the kind considered.   Second, sufficiently steep
i.e. large $n$ reduced density profiles exhibit an oblique inflection point,
where the threshold is found to be located at $n=n_{\rm th}=0.888715$.
Reduced pressure profiles, $\sP=p/\varpi=\theta^{n+1}$, $\varpi$ central
pressure, Lane-Emden fucntions, $\theta=(\rho/\lambda)^{1/n}$, and polytropic
curves, $\sP=\sP(\srho)$, are also plotted.   The
method can be extended to nonspherical polytropes with regard to a selected
direction, $\sr(\mu)=r(\mu)/R(\mu)=\xi(\mu)/\Xi(\mu)$.   The results can be
extended to polytropic spheres made of collisionless particles, for
polytropic index within a more restricted range, $1/2\le n\le5$.

\noindent

{\it keywords -
stars: equilibrium - galaxies: equilibrium - polytropic spheres.}
\end{small}
\end{quotation}

\section{Introduction} \label{intro}

Models in reduced variables are useful tools for the description of the
physical world, in that a single formulation relates to a whole set of
configurations.   For instance, let $\rho(r)$, $0\le r\le R$, be the density
profile of an assigned mass distribution along a selected direction, and let
$(r,\rho)=(r_s,\rho_s)$ be fixed nonzero scaling values.   Let reduced (or
scaled) variables be defined as $(\sr,\srho)=(r/r_s,\rho/\rho_s)$.
Accordingly, the reduced density profile reads $\srho(\sr)$, $0\le\sr\le\sR$,
which includes an infinity of density profiles, $r\to k_rr$, $k_r>0$,
$\rho\to k_\rho\rho$, $k_\rho>0$, in addition to the one under consideration.

A classical example of reduced variables can be found in van der Waals' theory
of real gases in connection with the critical point, lying on the critical
isothermal curve.   The coordinates of the critical point on the Clapeyron
plane are ${\sf P}_{\rm c}\equiv(V_{\rm c},p_{\rm c},T_{\rm c})$, where
$V_{\rm c}$ is the largest volume along the critical isothermal curve still
allowing a liquid phase, $p_{\rm c}$ is the lowest pressure along the critical
isothermal curve still allowing a liquid phase, and $T_{\rm c}$ is the
temperature along the critical isothermal curve i.e. the largest temperature
still allowing a liquid phase.

Though isothermal curves on the Clapeyron
plane, $({\sf O}Vp)$, are different for different gases, the contrary holds on
the reduced Clapeyron plane, $({\sf O}\sV\sP)$, where reduced isothermal
curves coincide for all gases as $\sV=V/V_{\rm c}$, $\sP=p/p_{\rm c}$,
$\sT=T/T_{\rm c}$, with extension to ideal gases.   In any case, the equation
of state reads $p=p(V,T)$.

A special case of astrophysical interest, $p=p(V)$ or $p=p(\rho)$, relates to
polytropic spheres or, in general, polytropes (e.g., Jeans 1929, Chap.\,IX,
\S\S235-239; Chandrasekhar 1939; Horedt 2004), which are self-gravitating
systems in hydrostatic equilibrium.   Related scaling radius and scaling
density are usually denoted as $r_s=\alpha$ and $\rho_s=\lambda$,
respectively, where $\alpha$ depends on the central pressure, the central
density, the density profile, and $\lambda$ is the central density.

In a widely investigated class of polytropic spheres, the reduced radial
coordinate is denoted as $\xi=r/\alpha$ and the reduced density as
$\theta^n=\rho/\lambda$, where $n$ is the polytropic index.   The cases of
astrophysical interest, $0\le n\le5$, range from null $(n=0)$ to infinite
$(n=5)$ degree of concentration or, in other words, from homogeneous to Roche
(e.g., Jeans 1929, Chap.\,IX, \S\S229-232) or Plummer (Plummer 1911) models,
according if the central density is divergent or finite, respectively.   The
reduced radius, $\Xi=R/\alpha$, is a monotonically increasing function of the
polytropic index, $n$, and $\Xi\to+\infty$ as $n\to5$ (Chandrasekhar 1939,
Chap.\,IV, \S4; Horedt 2004, Chap.\,2, \S2.5).   Accordingly, the reduced
density cannot be represented in a finite region of the reduced
$({\sf O}\xi\theta^n)$ plane for $0\le n\le5$.

To this aim, a different choice of reduced radial coordinates has to be
performed, $\sr=r/R=\xi/\Xi$, while the reduced density is left as
$\srho=\rho/\lambda=\theta^n$.   Under the restriction of null density on the
boundary, $\theta^n=0$, the whole set of reduced density profiles on the
$({\sf O}\sr\srho)$ plane lies within a square of unit sides parallel to the
coordinate axes, with a vertex on the origin.    A picture of the kind
considered could be useful in disclosing
additional features of reduced density profiles related to polytropic spheres.
To this subject, the current investigation is devoted.

The paper is organized as follows.   Reduced variables are introduced in
Section \ref{reva}, where the special case of ideal and real gases is
presented as a guidance example.   The special case of polytropic spheres is
considered in Section \ref{posp}, where reduced density profiles are plotted
on the reduced $({\sf O}\sr\srho)$ plane for several values of the polytropic
index, $0\le n\le5$, and the occurrence of an oblique inflection point is
studied in detail.   The reduced pressure profiles, the Lane-Emden functions,
and the polytropic curves, are similarly considered therein.  The discussion
and the concluding remarks are drawn in Section \ref{dico}.

\section{Reduced variables}
\label{reva}

Let the equilibrium configuration of a thermodynamical system be defined by
a set of physical parameters, $P_1$, $P_2$, ..., $P_N$.   Let $P_{s,1}$,
$P_{s,2}$, ..., $P_{s,N}$, be selected reference values or scaling parameters
with respect to the above mentioned ones.   Let the dimensionless parameters,
$\sPM_1$, $\sPM_2$, ..., $\sPM_N$, be defined as reduced or scaled parameters,
with respect to the above mentioned ones.   A description in terms of reduced
parameters includes all configurations where $P_i\to k_iP_i$, $k_i>0$; in
particular, $k_i=1$, $1\le i\le N$, relates to the system of interest.

As a guidance example, ideal and real gases shall be taken into consideration.
For further details and exhaustive presentation, an interested reader is
addressed to articles on the subject (e.g., Caimmi 2010, 2012) or specific
textbooks (e.g., Rostagni 1957; Landau and Lifchitz 1967).

The equation of state of ideal (e.g., Landau and Lifchitz 1967, Chap.\,IV,
\S42) and real (van der Waals 1873) gases, respectively, read:
\begin{lefteqnarray}
\label{eq:gid}
&& pV=kNT~~; \\
\label{eq:VdW}
&& \left(p+A\frac{N^2}{V^2}\right)(V-NB)=kNT~~;
\end{lefteqnarray}
where $p$ is the pressure, $V$ the volume, $T$ the temperature, $N$ the
particle number, $k$ the Boltzmann constant, $A$ and $B$ constants which
depend on the nature of the particles.   In particular, $B$ can be conceived
as the volume filled by a particle of real gas and the product, $NB$, as the
volume filled by all particles, or covolume (e.g., Landau and Lifchitz 1967,
Chap.\,IV, \S74).

Van der Waals' equation of state, Eq.\,(\ref{eq:VdW}), can be rewritten as
$p=p(V,T)$, and the partial derivatives, $(\partial p/\partial V)_{V,T}$,
$(\partial^2p/\partial V^2)_{V,T}$, can explicitly be expressed.   Van der
Waals isothermal curves may exhibit two extremum points (maximum and minimum),
allowing a liquid phase, or no extremum point, allowing no liquid phase.   The
threshold relates to the critical isothermal curve, which shows a single
extremum (horizontal inflection) point, where a liquid phase still occurs.
The above mentioned inflection point is defined as critical point and related
coordinates, ${\sf P}_{\rm c}\equiv(V_{\rm c},p_{\rm c},T_{\rm c})$, are
defined as critical volume, critical pressure, critical temperature,
respectively, where the last relates to the critical isothermal curve.

Owing to the mathematical properties of horizontal inflection points,
$(\partial p/\partial V)_{V_{\rm c},T_{\rm c}}=0$,
$(\partial^2p/\partial V^2)_{V_{\rm c},T_{\rm c}}=0$, which, together with
$p_{\rm c}=p(V_{\rm c},T_{\rm c})$, make a system of three equations in the
three unknowns, $V_{\rm c}$, $p_{\rm c}$, $T_{\rm c}$.   The solution is
(e.g., Rostagni 1957, Chap.\,XII, \S20; Landau and Lifchitz 1967, Chap.\,VIII,
\S85; Caimmi 2010, 2012):
\begin{lefteqnarray}
\label{eq:Vc}
&& V_{\rm c}=3NB~~; \\
\label{eq:Tc}
&& T_{\rm c}=\frac8{27}\frac AB\frac1k~~; \\
\label{eq:pc}
&& p_{\rm c}=\frac1{27}\frac A{B^2}~~;
\end{lefteqnarray}
in terms of the covolume, $NB$, and the constants, $A$, $B$, $k$.

With regard to the reduced variables:
\begin{equation}
\label{eq:rv}
\sV=\frac V{V_{\rm c}}~~;\qquad\sP=\frac p{p_{\rm c}}~~;
\qquad\sT=\frac T{T_{\rm c}}~~;
\end{equation}
the ideal gas equation of state, Eq.\,(\ref{eq:gid}),
and van der Waals' equation of state, Eq.\,(\ref{eq:VdW}),
take the expression:
\begin{lefteqnarray}
\label{eq:ri}
&& \sP\sV=\frac83\sT~~; \\
\label{eq:rW1}
&& \left(\sP+\frac3{\sV^2}\right)\left(\sV-\frac13\right)=\frac83\sT~~;
\end{lefteqnarray}
where the domain is $\sV>0$, $\sV>1/3$, respectively.

\begin{figure*}[t]  
\begin{center}      
\includegraphics[scale=0.8]{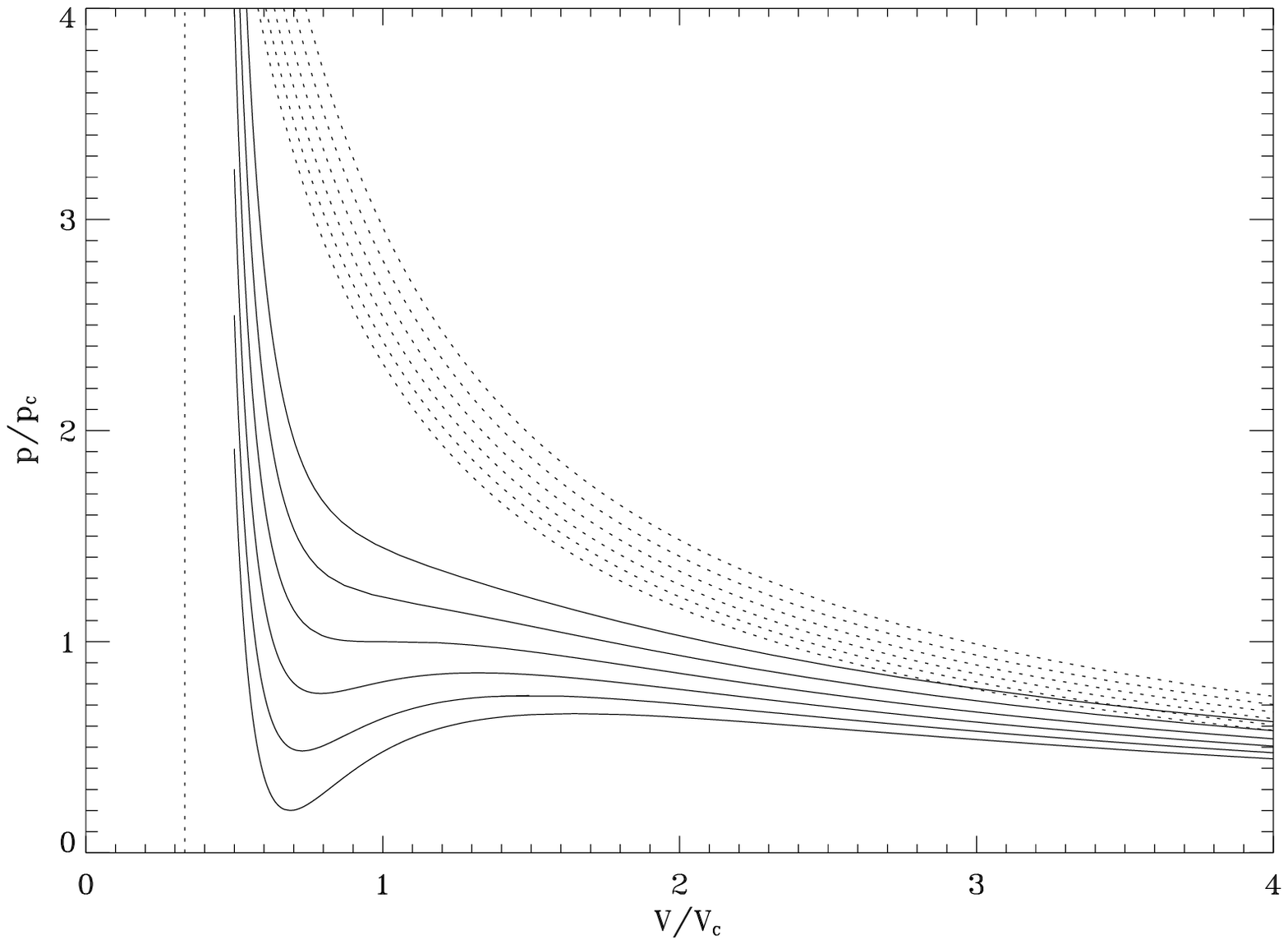}                      
\caption[ddbb]{Isothermal curves in reduced variables for ideal (dotted) and
van der Waals' (full) gases, respectively.   The reduced temperature on each
curve (from bottom to top in both cases) is $T/T_{\rm c}=20/23,$ 20/22, 20/21,
20/20, 20/19, 20/18, respectively.   The horizontal asymptote is the
horizontal axis.   The vertical asymptote of ideal
isothermal curves is the vertical axis.   The vertical asymptote of van der
Waals' isothermal curves is shown as a dotted line, which defines the reduced
covolume, $V/V_{\rm c}=1/3$.   No extremum point exists above the critical
isothermal curve, $T/T_{\rm c}=1$.   See text for further details.}
\label{f:gaspv}     
\end{center}       
\end{figure*}

It is worth emphasyzing Eqs.\,(\ref{eq:ri})-(\ref{eq:rW1}) are independent of
the nature of the gas, contrary to Eqs.\,(\ref{eq:gid})-(\ref{eq:VdW}), hence
the great advantage of reduced variables with respect to physical variables.
Reduced isothermal curves exhibit a horizontal asymptote, $\sP=0$, and a
vertical asymptote, $\sV=0$ and $\sV=1/3$ for ideal and real gases,
respectively.   The reduced critical isothermal curve together a few
neighbourhing ones, related to $\sT=20/23$, 20/22, 20/21, 20/20, 20/19, 20/18,
are shown in Fig.\,\ref{f:gaspv} (full curves) together with their
counterparts for ideal gases (dotted curves).   The critical point is
${\sf P}_{\rm c}\equiv(1,1,1)$.

\section{Polytropic spheres}
\label{posp}
\subsection{General considerations}
\label{geco}

Polytropes are special cases of barotropes i.e. self-gravitating fluids in
hydrostatic equilibrium where the equation of state reads $p=p(\rho)$ or,
restricting to polytropes (e.g., Jeans 1929, Chap.\,IX, \S\S235-239):
\begin{lefteqnarray}
\label{eq:pirho}
&& p=K\left(\rho^{1+1/n}-\rho_{\rm b}^{1+1/n}\right)~~;
\end{lefteqnarray}
where $K$ is a constant, $n$ the polytropic index and $\rho_{\rm b}$ the
density on the boundary, which is usually taken equal to zero (e.g.,
Chandrasekhar 1939, Chap.\,4; Horedt 2004, Chap.\,2).   The condition of
hydrostatic equilibrium via Poisson equation reads (e.g., Caimmi 1980):
\begin{lefteqnarray}
\label{eq:Pois}
&& \Delta{\cal V}_{\rm G}=-4\pi G\rho~~; \\
\label{eq:hydr}
&& {\cal V}_{\rm G}=K(n+1)\rho^{1/n}+{\cal V}_{\rm b}~~;
\end{lefteqnarray}
where ${\cal V}_{\rm G}$ is the gravitational potential and ${\cal V}_{\rm b}$
a normalization constant.   From this point on, it shall be assumed
$\rho_{\rm b}=0$ for simplicity.

Let $\lambda$ be the central density and $\alpha$ the scaling radius, defined
as:
\begin{lefteqnarray}
\label{eq:alfa}
&& \alpha=\left[\frac{(n+1)\varpi}{4\pi G\lambda^2}\right]^{1/2}=
\left[\frac{(n+1)K\lambda^{1/n}}{4\pi G\lambda}\right]^{1/2}~~;
\end{lefteqnarray}
where $\varpi$ is the central pressure and $K\lambda^{1+1/n}$ is dimensioned
as a pressure.

With regard to the reduced variables:
\begin{lefteqnarray}
\label{eq:csi}
&& \xi=\frac r\alpha~~;\qquad0\le r\le R~~; \\
\label{eq:ten}
&& \srho=\theta^n(\xi)=\frac{\rho(r)}\lambda~~;\qquad0\le\xi\le\Xi~~;
\end{lefteqnarray}
where $\Xi=R/\alpha$ is the reduced radius, the substitution of
Eqs.\,(\ref{eq:hydr})-(\ref{eq:ten}) into (\ref{eq:Pois}) after some algebra
yields the Lane-Emden equation (e.g., Chandrasekhar 1939, Chap.\,4, \S2;
Caimmi 1980; Horedt 2004, Chap.\,2, \S2.1):
\begin{lefteqnarray}
\label{eq:LE}
&& \theta^\pprime+\frac2\xi\theta^\prime=-\theta^n~~;
\end{lefteqnarray}
where the prime denotes derivation with respect to $\xi$.

Density profiles
of astrophysical interest (i.e. decreasing with increasing radial coordinate)
correspond to the range of polytropic index, $0\le n\le5$, where $n=0$ relates
to homogeneous models and $n=5$ to Roche (e.g., Jeans 1929, Chap.\,IX,
\S\S229-232) or Plummer (Plummer 1911) models, according if the central
density, $\lambda$, is divergent or finite, respectively.   Reduced radii are
monotonically increasing from $\Xi=\sqrt6~(n=0)$ to $\Xi\to+\infty~(n=5)$.

The divergence of the reduced radius as $n\to5$ makes a representation of
reduced density profiles on the $({\sf O}\xi\srho)$ plane of little
utility, in that interesting features could be lost.   As $0\le\theta^n\le1$,
a reduced radial coordinate within a similar range would be needed.   To this
respect, let an additional reduced radial coordinate be defined as:
\begin{lefteqnarray}
\label{eq:rcs}
&& \sr=\frac rR=\frac\xi\Xi~~;\qquad0\le\sr\le1~~;
\end{lefteqnarray}
and let the reduced density profile, $\srho=\srho(\sr)$, be
considered and plotted on the $({\sf O}\sr\srho)$ plane, which implies the
knowledge of $\theta^n(\xi/\Xi)$, $\Xi$, for selected values of $n$.

\subsection{Source of data}
\label{soda}

Physical parameters for sequences of rigidly rotating polytropes were
determined in an earlier investigation (Caimmi 1983, hereafter quoted as C83),
from nonrotating to maximally rotating (i.e. up to centrifugal support on the
equatorial plane) configurations.   Unfortunately, computations cannot be
repeated as the original computer code is still in cards and no conversion
into electronic format was tried in the past.   For this reason, data used in
the current paper are taken as specified below.
\begin{description}
\item[$\bullet$\hspace{6mm}]
For polytropic indexes, $n=0,1,5$, where density profiles can be expressed
analytically, standard formulae are used (e.g., Caimmi 1980).
\item[$\bullet$\hspace{6mm}]
For integer and half-integer polytropic indexes, with the addition of
$n=4.99$, seven-digit tables of Lane-Emden functions (Horedt 1986, hereafter
quoted as H86) are used.
\item[$\bullet$\hspace{6mm}]
For quarter-integer polytropic indexes, with the addition of $n=0.001$, 0.010,
0.050, 0.100, 0.200, 0.808, a computer code with starting series solution
followed by a fourth-order Runge-Kutta interpolation method, when an assigned
tolerance is exceeded, is used.
\end{description}
Accordingly, reduced density profiles can be plotted on the
$({\sf O}\sr\srho)$ plane.

\subsection{Results}
\label{resu}

Plotting in terms of the reduced radius, via Eq.\,(\ref{eq:rcs}) implies the
knowledge of the scaled radius, $\Xi$.   Related values from the sources
mentioned above are listed in Table \ref{t:spol} for several
\begin{table*}
\caption[par]{The scaled radius, $\Xi$, of polytropic spheres for polytropic
index, $n$, within the range, $0\le n\le5$, according to the present paper
(pp) and earlier investigations (C83; H86).   Cases related to symbols (s) are
plotted in Fig.\,\ref{f:poli}.   See text for further details.}
\label{t:spol}
\begin{center}
\begin{tabular}{llllc}
\hline
\multicolumn{1}{c}{$n$} &
\multicolumn{1}{c}{$\Xi~$(pp)} &
\multicolumn{1}{c}{$\Xi~$(C83)} &
\multicolumn{1}{c}{$\Xi~$(H86)} &
\multicolumn{1}{c}{s} \\
\hline\noalign{\smallskip}
0.000 &               & 2.44948974E+00 & 2.44948974E+00 &   f         \\
0.001 & 2.4499580E+00 &                &                &             \\
0.010 & 2.4547382E+00 & 2.45488185E+00 &                &             \\
0.050 & 2.4740123E+00 & 2.47669981E+00 &                &             \\
0.100 & 2.4977252E+00 & 2.50454496E+00 &                &   d         \\
0.200 & 2.5510944E+00 & 2.56221918E+00 &                &             \\
0.250 & 2.5797948E+00 & 2.59208980E+00 &                & $\ast$      \\
0.500 & 2.7404437E+00 & 2.75269805E+00 & 2.75269805E+00 & $\diamond$  \\
0.750 & 2.9345165E+00 & 2.93451648E+00 &                &             \\
0.808 & 2.9801387E+00 & 2.98013932E+00 &                & +           \\
1.000 & 3.1415925E+00 & 3.14159265E+00 & 3.14159265E+00 &   f         \\
1.250 & 3.3791024E+00 & 3.37910200E+00 &                &             \\
1.500 & 3.6537544E+00 & 3.65375374E+00 & 3.65375374E+00 & $\triangle$ \\
1.750 & 3.9743875E+00 & 3.97438776E+00 &                &             \\
2.000 & 4.3528750E+00 & 4.35287460E+00 & 4.35287460E+00 & $\times$    \\
2.250 & 4.8055144E+00 & 4.80551285E+00 &                &             \\
2.500 & 5.3552750E+00 & 5.35527546E+00 & 5.35527546E+00 & $\quadr$    \\
2.750 & 6.0355700E+00 & 6.03557001E+00 &                &             \\
3.000 & 6.8968500E+00 & 6.89684862E+00 & 6.89684862E+00 & $\ast$      \\
3.250 & 8.0189350E+00 & 8.01893753E+00 &                &             \\
3.500 & 9.5358000E+00 & 9.53580534E+00 & 9.53580534E+00 & $\diamond$  \\
3.750 & 1.1690285E+01 & 1.16902937E+01 &                &             \\
4.000 & 1.4971525E+01 & 1.49715463E+01 & 1.49715463E+01 & +           \\
4.250 & 2.0529055E+01 & 2.05291013E+01 &                &             \\
4.500 & 3.1836310E+01 & 3.18364632E+01 & 3.18364632E+01 & $\triangle$ \\
4.750 & 6.6386250E+01 & 6.63870957E+01 &                &             \\
4.850 & 1.1295301E+02 &                &                &   d         \\
4.990 &               &                & 1.75818915E+03 & $\times$    \\
5.000 &               & $+\infty$      & $+\infty$      &   f         \\
\noalign{\smallskip}
\hline
\end{tabular}
\end{center}
\end{table*}
polytropic indexes, $n$, with regard to the computer code used in the present
paper (pp), results from the parent paper (C83) published later (Caimmi 1985),
and results from seven-digit tables of Lane-Emden functions (H86).   The
computer code does not hold for $n=0$ due to the occurrence of undetermined
forms of the kind, 0/0 and so on, and for $n=5$ due to memory overflow.  An
inspection of Table \ref{t:spol} discloses that, within
$\theta(\Xi)<5\cdot10^{-7}$, $\Xi$ agrees with its counterpart from C83 and/or H86
within a few percent for $0<n<0.5$ and within $10^{-4}$ or less for
$0.75\le n\le4.75$.

Concerning reduced density profiles not included in the seven-digit tables of
Lane-Emden functions (H86),
computed scaled radii, $\Xi$, are used in determining the reduced radial
coordinate, $\sr=r/R=\xi/\Xi$, for the following reason.   Within the range of
interest, $0\le\xi\le\Xi$, the Lane-Emden function, $\theta(\xi)$,
$1\ge\theta(\xi)\ge0$, is monotonically decreasing, then
overstimated/understimated $\Xi$ implies overstimated/understimated $\xi$ for
fixed $\theta$, with respect to related true values listed in Table
\ref{t:spol}, or in other words
$\Delta\xi/\Delta\Xi>0$, where $\Delta\xi$, $\Delta\Xi$, are computation
errors.   Accordingly, the computed reduced radial coordinate reads
$\sr=(\xi+\Delta\xi)/(\Xi+\Delta\Xi)$, which is closer to the true value,
$\sr=\xi/\Xi$, than $\sr=(\xi+\Delta\xi)/\Xi$.

\begin{figure*}[t]  
\begin{center}      
\includegraphics[scale=0.8]{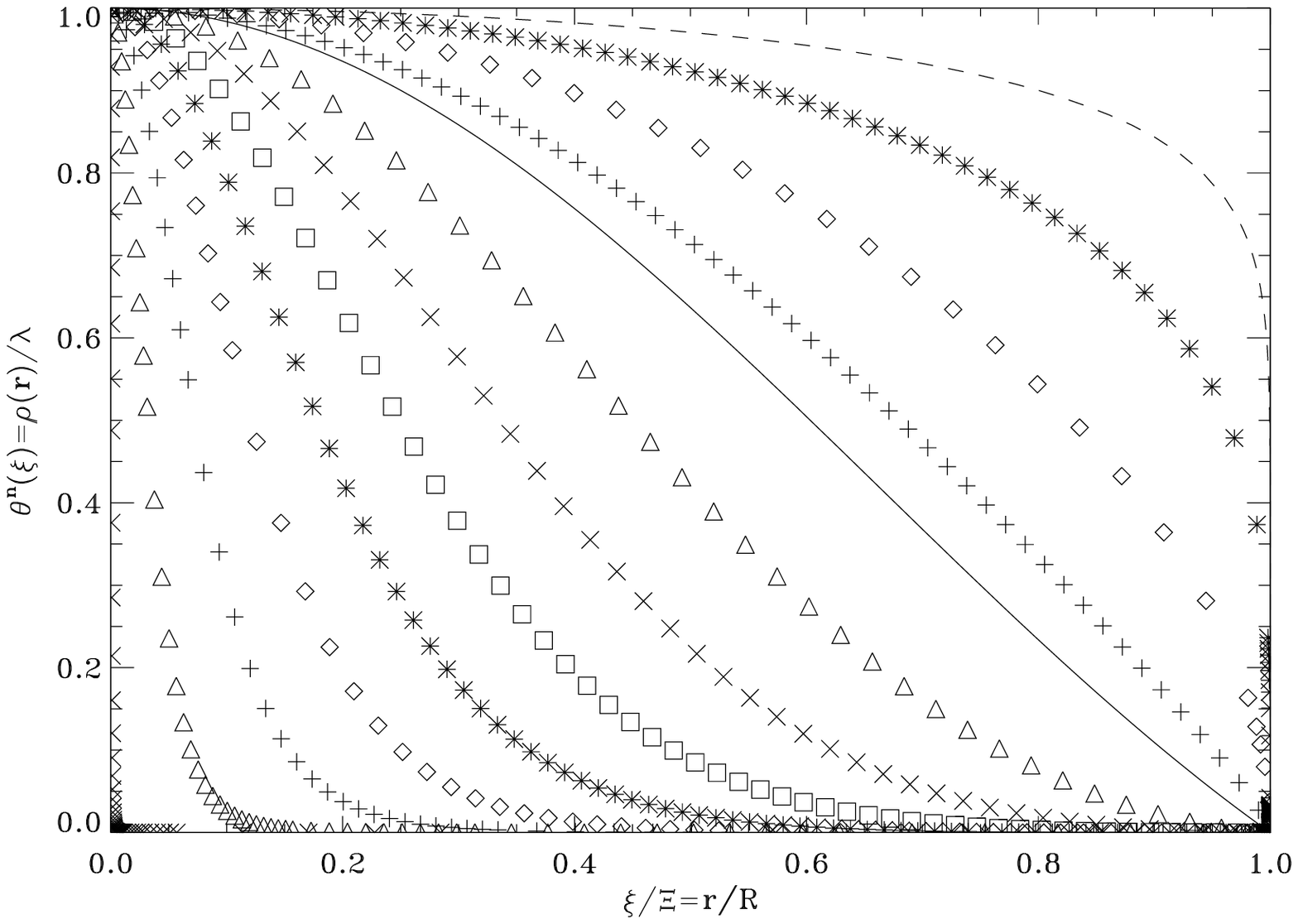}                      
\caption[ddbb]{Density profiles of polytropic spheres in reduced variables,
$\theta^n=\rho(r)/\lambda$ vs. $\xi/\Xi=r/R$, for different values of
polytropic index, $0\le n\le5$, as listed in Table \ref{t:spol}, where the
corresponding symbol is also shown.
Full curves (f) relate to $n=0$ (top and
right side of the box), $n=5$ (left and bottom side of the box), and $n=1$,
for which density profiles can be expressed analytically.   The dashed curve
(d) relates to $n=0.1$.   Symbols upside with respect to the full curve
$(n=1)$ correspond to $n=0.25$ (asterisks), 0.5 (diamonds), 0.808 (crosses),
starting from top right.   Symbols downside with respect to the full curve
correspond to $n=1.5$ (triangles), 2.0 (saltires), 2.5 (squares), 3.0
(asterisks), 3.5 (diamonds), 4.0 (crosses), 4.5 (triangles), 4.99 (saltires),
starting from the full curve towards bottom left.  Source of data: $n=0.5$ and
$1<n<5$ (H86); $n=0.1$, 0.25, 0.808, (present paper).   See text for
further details.}
\label{f:poli}     
\end{center}       
\end{figure*}

Reduced density profiles, $\srho=\rho(r)/\lambda=\theta^n(\xi)$, vs. reduced
radial coordinates, $\sr=r/R=\xi/\Xi$, for polytropic index within the range,
$0\le n\le5$, are plotted in Fig.\,\ref{f:poli} where symbol captions are also
listed in Table \ref{t:spol}.   Full $(n=0,1,5)$ and dashed $(n=0.1)$ curves
with the addition of symbols $(n=0.25, 0.808)$ are related to exact and
computed (pp) solutions of the Lane-Emden equation, respectively.   Remaining
symbols (integer and half-integer $n$, $1<n<5$, with the addition of $n=0.50,
4.99$) are from seven-digit tables of Lane-Emden function (H86).

The limiting case, $n=0$, is represented by the top and right side of the box
in Fig.\,\ref{f:poli}.   To this respect, it is worth emphasyzing polytropic
spheres with $n=0$ are conceptually different from MacLaurin spheres.   More
specifically, the latter are incompressible while the former are compressible
but with infinite pressure inside and null pressure on the boundary conformly
to hydrostatic equilibrium.   Then the density on the boundary of polytropic
spheres is null while it remains finite on the boundary of MacLaurin spheres,
where a discontinuity arises.

The limiting case, $n=5$, is represented by the left and bottom side of the
box in Fig.\,\ref{f:poli}, due to $\Xi\to+\infty$, which implies finite $\xi$
on the origin and infinite $\xi$ for the remaining of the domain,
$0<\xi\le\Xi$.   In other words, the massive body is ``compressed'' into the
origin while the vanishing atmosphere extends up to $\xi/\Xi=1$, similarly to
Roche models in the physical space (e.g., Jeans 1929, Chap.\,IX, \S\S229-232.

Concerning the remaining cases, an inspection of Fig.\,\ref{f:poli} shows
reduced density profiles can be divided into two main classes, namely
exhibiting one or no oblique inflection point for $n\ge1$ and $n\le0.808$,
respectively, with the threshold lying in between.   The monotonic trend of
the reduced density, $\srho=\theta^n(\xi/\Xi)$, implies no extremum point and,
in turn, oblique inflection points related to $\diff^2\srho/\diff\sr^2=0$,
which via Eq.\,(\ref{eq:ten}) after little algebra can be expressed as:
\begin{equation}
\label{eq:flob}
n\Xi^2\theta^{n-2}[(n-1)(\theta^\prime)^2+\theta\theta^\pprime]=0~~;
\end{equation}
where the prime denotes derivation with respect to $\xi$.
The substitution of Eq.\,(\ref{eq:LE}) into (\ref{eq:flob}) yields:
\begin{equation}
\label{eq:cfob}
(n-1)(\theta^\prime)^2-\frac2\xi\theta\theta^\prime-\theta^{n+1}=0~~;
\end{equation}
which is the condition for the existence of an oblique inflection point in the
case under discussion.

The lowest $n$ for which Eq.\,(\ref{eq:cfob}) is still satisfied is found
numerically using the computer code described above, and the result is
$n=0.888715$ where $\Xi=3.0459945$ and the inflection point occurs at
$\xi=2.650000$ within a tolerance
$\epsilon=5\cdot10^{-7}$.   Related reduced density profile is not plotted in
Fig.\,\ref{f:poli} to avoid confusion.

\begin{figure*}[t]  
\begin{center}      
\includegraphics[scale=0.8]{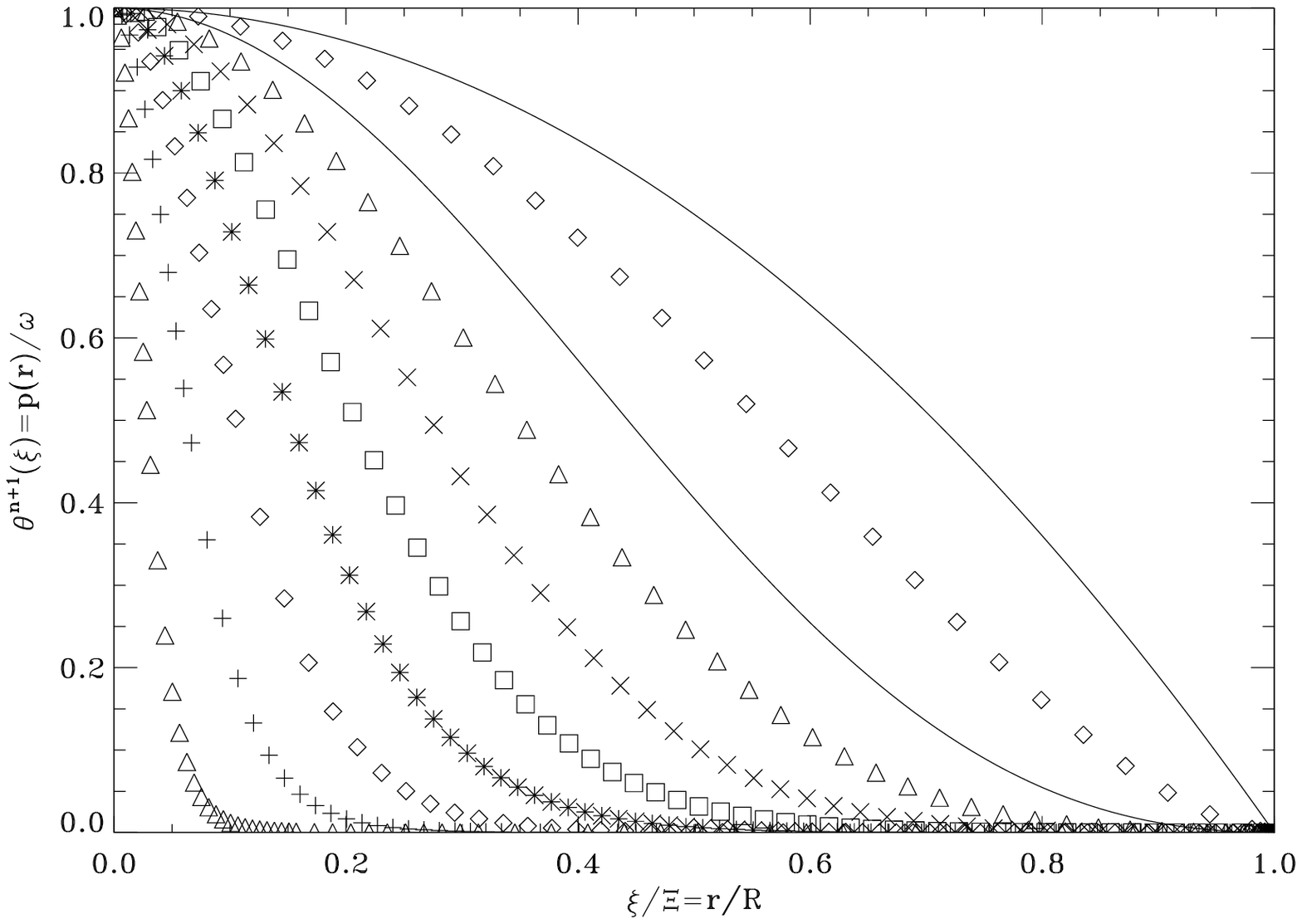}                      
\caption[ddbb]{Pressure profiles of polytropic spheres in reduced variables,
$\theta^{n+1}=p(r)/\varpi$ vs. $\xi/\Xi=r/R$, for integer and half-integer
values of polytropic index, $0\le n\le5$, as listed in Table \ref{t:spol},
where the corresponding symbol is also shown.
Full curves relate to $n=0$ (top right), $n=5$ (left and bottom side of
the box), and $n=1$, for which pressure profiles can be expressed
analytically.   Symbols correspond to $n=0.5$ (diamonds),
$n=1.5$ (triangles), 2.0 (saltires), 2.5 (squares), 3.0
(asterisks), 3.5 (diamonds), 4.0 (crosses), 4.5 (triangles), 
starting from the top right towards bottom left.   Data are from seven-digit
tables of Lane-Emden functions (H86).   See text for further details.}
\label{f:polp}     
\end{center}       
\end{figure*}

Reduced pressure profiles, $\sP=p(r)/\varpi=\theta^{n+1}(\xi)$, vs. reduced
radial coordinates, $\sr=r/R=\xi/\Xi$, for integer and half-integer polytropic
index within the range,
$0\le n\le5$, are plotted in Fig.\,\ref{f:polp} where symbol captions are as
in Fig.\,\ref{f:poli} and data are from seven-digit tables of Lane-Emden
function (H86).
The limiting case, $n=5$, is represented by the left and bottom side of the
box in Fig.\,\ref{f:polp}.

An inspection of Fig.\,\ref{f:polp} shows
reduced pressure profiles can be divided into two main classes, namely
exhibiting one or no oblique inflection point for $n\ge0.5$ and $n<0.5$,
respectively.   The monotonic trend of
the reduced pressure, $\sP=\theta^{n+1}(\xi/\Xi)$, implies no extremum point
and, in turn, oblique inflection points related to $\diff^2\sP/\diff\sr^2=0$,
which via Eq.\,(\ref{eq:ten}) after little algebra can be expressed as:
\begin{equation}
\label{eq:flop}
(n+1)\Xi^2\theta^{n-1}[n(\theta^\prime)^2+\theta\theta^\pprime]=0~~;
\end{equation}
where the prime denotes derivation with respect to $\xi$.
The substitution of Eq.\,(\ref{eq:LE}) into (\ref{eq:flop}) yields:
\begin{equation}
\label{eq:cfop}
n(\theta^\prime)^2-\frac2\xi\theta\theta^\prime-\theta^{n+1}=0~~;
\end{equation}
which is the condition for the existence of an oblique inflection point in the
case under discussion.

In the special case, $n=0$, the Lane-Emden function reads (e.g., Caimmi 1980)
$\theta(\xi)=1-\xi^2/6$, hence $\theta^\prime(\xi)=-\xi/3$ and
Eq.\,(\ref{eq:cfop}) reduces to:
\begin{equation}
\label{eq:cf0p}
\theta^n=\frac23~~;\qquad n=0~~;
\end{equation}
which holds on the boundary, $\xi=\Xi=\sqrt6$, keeping in mind the pressure,
and then the density, has to be null on the boundary.   Then a ``vertical''
inflection point of the reduced pressure profile takes place on the boundary.
Accordingly, all reduced pressure profiles,
$\sP=p(r)/\varpi=\theta^{n+1}(\xi)$, within the
range, $0\le n\le5$, exhibit an oblique inflection point.

\begin{figure*}[t]  
\begin{center}      
\includegraphics[scale=0.8]{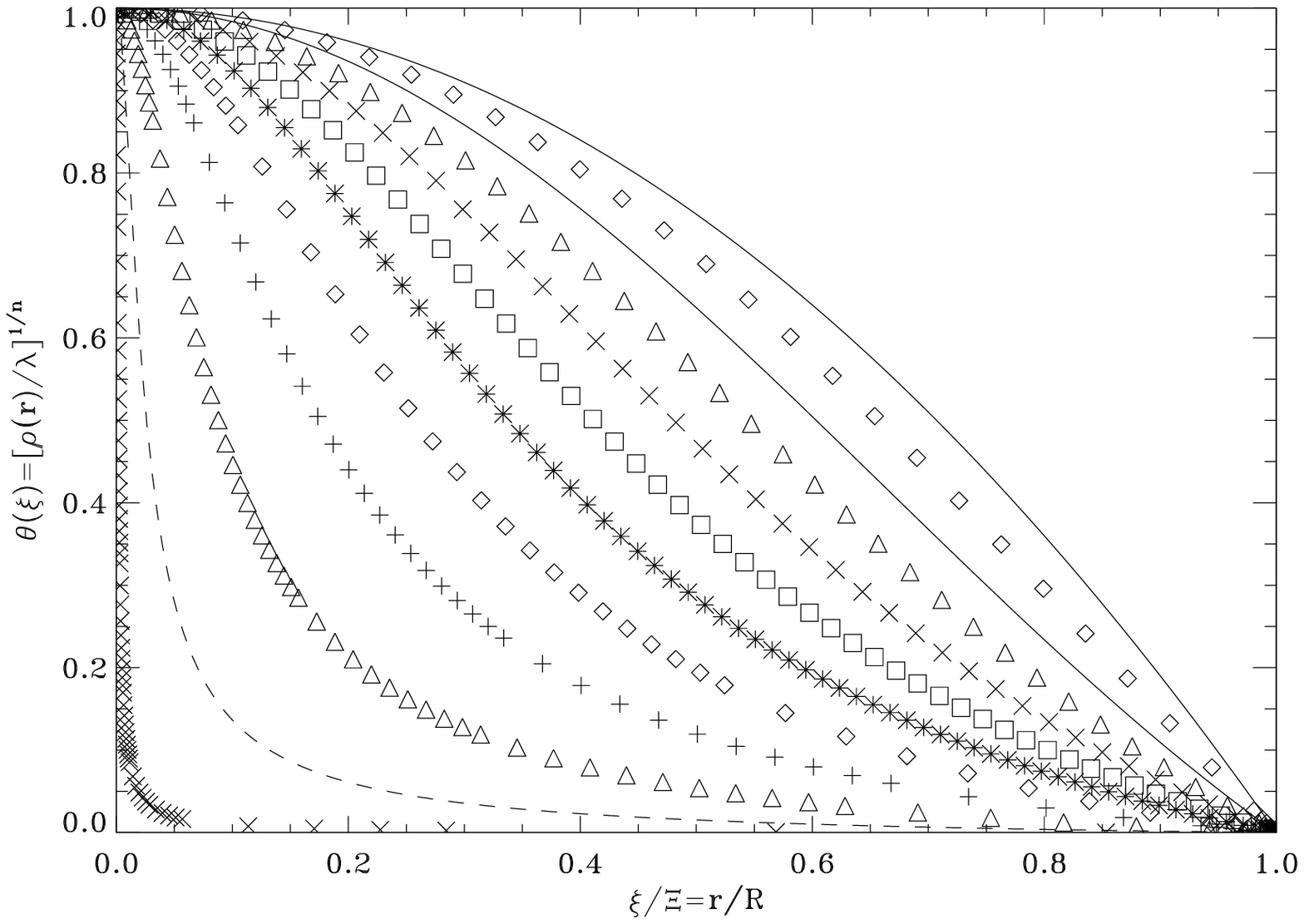}                      
\caption[ddbb]{The Lane-Emden function in reduced variables,
$\theta=[\rho(r)/\lambda]^{1/n}$ vs. $\xi/\Xi=r/R$, for integer and
half-integer values of polytropic index, $0\le n\le5$, with the addition of
$n=4.85, 4.99,$ as listed in Table \ref{t:spol},
where the corresponding symbol is also shown.
Full curves relate to $n=0$ (top right), $n=5$ (left and bottom side of
the box), and $n=1$, for which the Lane-Emden function can be expressed
analytically.   Symbols correspond to $n=0.5$ (diamonds),
$n=1.5$ (triangles), 2.0 (saltires), 2.5 (squares), 3.0
(asterisks), 3.5 (diamonds), 4.0 (crosses), 4.5 (triangles), 4.85 (dashed),
4.99 (saltires),
starting from top right towards bottom left.   Data are from seven-digit
tables of Lane-Emden functions (H86) except for $n=4.85$ (pp).   See text for
further details.}
\label{f:polf}     
\end{center}       
\end{figure*}

The Lane-Emden functions, $\theta=[\rho(r)/\lambda]^{1/n}$, vs. reduced
radial coordinates, $\sr=r/R=\xi/\Xi$, for integer and half-integer polytropic
index within the range, $0\le n\le5$, with the addition of $n=4.85, 4.99$, are
plotted in Fig.\,\ref{f:polf} where symbol captions are as
in Fig.\,\ref{f:poli} and data are from seven-digit tables of Lane-Emden
function (H86) except for the case, $n=4.85$, where computations were
performed as outlined above.
The limiting case, $n=5$, is represented by the left and bottom side of the
box in Fig.\,\ref{f:polf}.

An inspection of Fig.\,\ref{f:polf} shows the Lane-Emden function is
characterized by the occurrence of an oblique inflection point, from the
boundary $(n=0)$ to the centre $(n=5)$.   In fact, the monotonic trend of
the Lane-Emden function, $\theta(\xi/\Xi)$, implies no extremum point
and, in turn, oblique inflection points related to
$\diff^2\theta/\diff\sr^2=0$, or $\Xi^2\theta^\pprime=0$, which via
Eq.\,(\ref{eq:LE}) can be expressed as:
\begin{equation}
\label{eq:cf0f}
\frac2\xi\theta^\prime=-\theta^n~~;
\end{equation}
that in the special case, $n=0$, reduces to Eq.\,(\ref{eq:cf0p}).   Then a
``vertical'' inflection point of the Lane-Emden function takes place
on the boundary.   Accordingly, all Lane-Emden functions,
$\theta=[\rho(r)/\lambda]^{1/n}$, within the
range, $0\le n\le5$, exhibit an oblique inflection point.

\begin{figure*}[t]  
\begin{center}      
\includegraphics[scale=0.8]{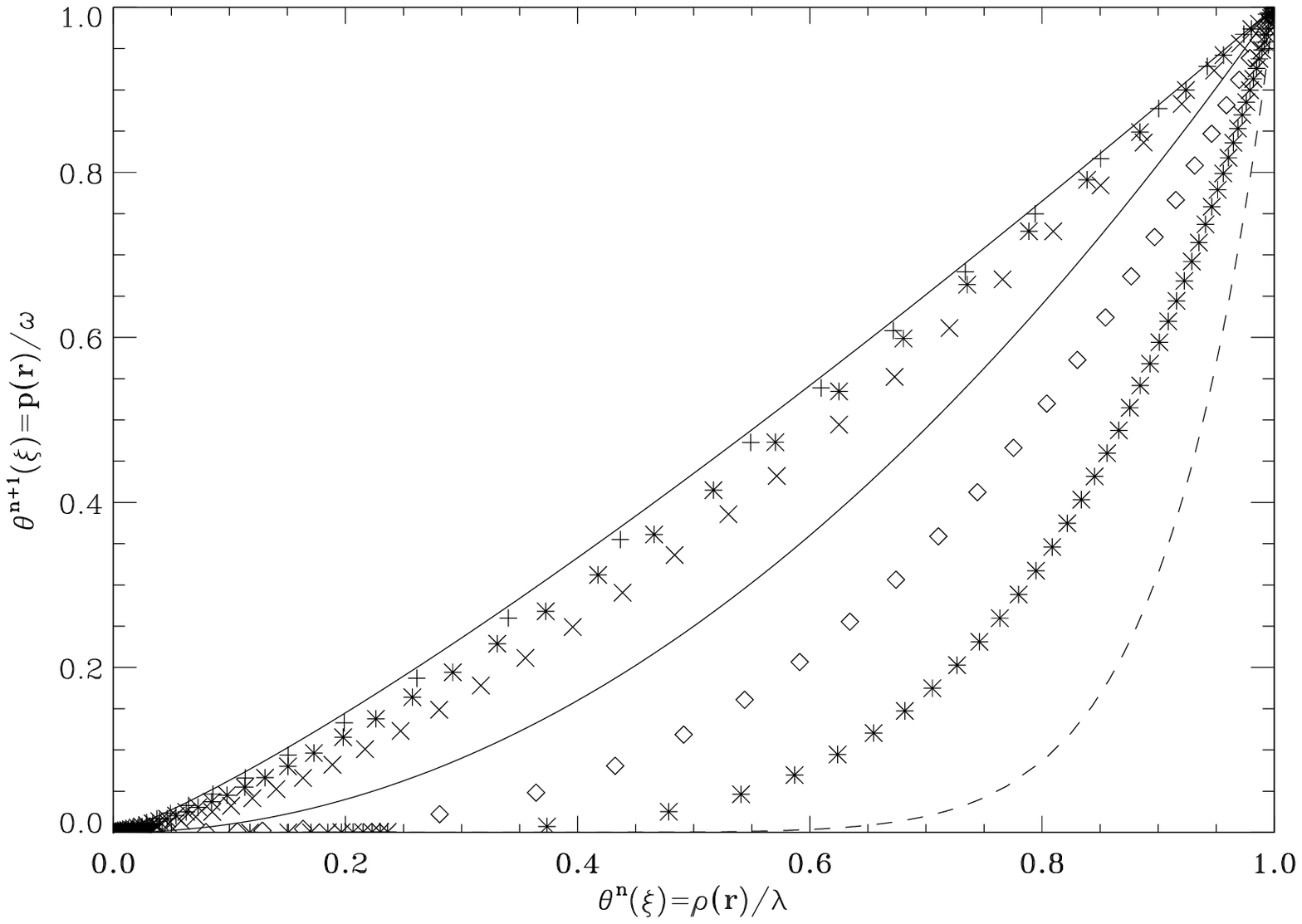}                      
\caption[ddbb]{Reduced polytropic curves, $\sP=\sP(\srho)$ or $\theta^{n+1}$
vs. $\theta^n$, for integer values of polytropic index, $0\le n\le5$, with the
addition of $n=0.50, 0.25, 0.10,$ as listed in Table \ref{t:spol},
where the corresponding symbol is also shown.
Full curves relate to $n=0$ (right and bottom side of the box), $n=5$ (top
left), and $n=1$, for which the Lane-Emden function can be expressed
analytically.   The dashed curve relates to $n=0.1$.   Symbols correspond to
$n=0.25$ (asterisks), 0.5 (diamonds), 2.0 (saltires), 3.0 (asterisks), 4.0
(crosses), starting from bottom right towards top left.   Data are from
seven-digit tables of Lane-Emden functions (H86) except for $n=0.10, 0.25$
(pp).   The centre and the boundary of the sphere correspond to $(1,1)$ and
$(0,0)$, respectively.   See text for further details.}
\label{f:pols}     
\end{center}       
\end{figure*}

The reduced polytropic curves, $\sP=\sP(\srho)$ or $\theta^{n+1}$ vs.
$\theta^n$, for integer
polytropic index within the range, $0\le n\le5$, with the addition of
$n=0.50, 0.25, 0.10$, are plotted in Fig.\,\ref{f:pols} where symbol captions
are as in Fig.\,\ref{f:poli} and data are from seven-digit tables of
Lane-Emden function (H86) except for the cases $n=0.25, 0.10$, where
computations were performed as outlined above.
The centre and the boundary of the sphere correspond to $(1,1)$ and $(0,0)$,
respectively.
The limiting case, $n=0$, is represented by the right and bottom side of the
box in Fig.\,\ref{f:pols}.

\section{Discussion and conclusion}
\label{dico}

An application to polytropic spheres has shown the usefulness and the power of
models in reduced variables.   With regard to the plane, $({\sf O}\sr\srho)$,
reduced density profiles for polytropic indexes, $0\le n\le5$, completely fill
a square of unit side, $0\le\sr\le1$, $1\ge\srho\ge0$, where the top and the
right side relate to homogeneous models $(n=0)$, while the left and the bottom
side relate to Roche and Plummer models $(n=5)$.   The last case can be
related to extremely inhomogeneous mass distributions, where the extension is
finite and the density is
nonzero only on the centre for Roche models (e.g., Jeans 1929, Chap.\,IX,
\S\S229-232), while the extension is infinite and the density is nonzero
provided the distance from the centre remains finite for Plummer models
(Plummer 1911).

Polytropic spheres can be conceived as matter distributions where the reduced
slope, $\kappa=\diff\srho/\diff\sr$, lies between the extreme limits,
$\kappa=0$ $(n=0)$ and $\kappa\to-\infty$ $(n=5)$.   From a geometrical point
of view,
the transition of reduced density profiles towards $\kappa\to0$ or $n\to0$
appears similar to the transition of Fermi-Dirac distribution functions
towards zero absolute temperature, $T\to0$ (e.g., Landau and Lifchitz 1967,
Chap.\,V, \S56).

The reduced slope, via Eqs.\,(\ref{eq:ten}) and (\ref{eq:rcs}) takes the
explicit form:
\begin{lefteqnarray}
\label{eq:res}
&& \kappa(\xi)=\frac{\diff\srho}{\diff\sr}=n\Xi\theta^{n-1}\theta^\prime~~;
\end{lefteqnarray}
where $\theta^\prime$ remains finite.   Accordingly, $\kappa(\Xi)\to-\infty$
for $0\le n<1$, $\kappa(\Xi)=1/\pi$ for $n=1$, $\kappa(\Xi)=0$ for $1<n\le5$.

The occurrence of an oblique inflection point on reduced density profiles for
sufficiently large $n$, with the threshold at $n=n_{\rm th}=0.888715$,
suggests a definition of ``steep'' and ``mild'' reduced density profiles as
related to $n\ge n_{\rm th}$ and $n<n_{\rm th}$, respectively, with regard to
polytropic spheres.   In addition, mild density profiles could be related to
isothermal curves of real gases above the critical one, where no inflection
point appears, and steep density profiles could be related to isothermal
curves of real gases below the critical one, where (two extremum points and
then) two inflection points appear, as shown in Fig.\,\ref{f:gaspv}.

The results of the current paper can be extended to polytropic spheres made of
collisionless particles, keeping in mind collisionless polytropes in rigid
rotation have an exact collisional counterpart within the range of polytropic
index, $1/2\le n\le5$ (Vandervoort 1980).

The results can also be extended to nonspherical polytropes, provided reduced
density profiles are considered along a selected direction, $r=r(\mu)$, hence
$\sr(\mu)=r(\mu)/R(\mu)=\xi(\mu)/\Xi(\mu)$, while
$\srho=\rho(r)/\lambda=\theta^n(\xi)$ remains unchanged in connection with
isopycnic surfaces, $r=r(\mu)$ or $\xi=\xi(\mu)$.

In conclusion, reduced variables appear to be not restricted to the Clapeyron
plane as initially conceived (van der Waals 1873), but they can be
successfully extended to other physical situations as shown for polytropic
spheres, which provides additional credit to van der Waals' original work.

\section*{Acknowledgements}

The author is deeply indebted to G.P. Horedt for making available his FORTRAN
program%
\footnote{
The program could not be used due to the lack of FORTRAN compiler in
the author's computer.   Calculations were performed using a GWBASIC program
of lower but still acceptable precision.
}
(H86) and tables of Lane-Emden functions (Horedt 2004) in TEX format.

\end{document}